\documentclass{ws-procs9x6}
\usepackage{epsfig}
\usepackage{wrapft}
\begin{document}

\title {QCD Thermodynamics:\\
 Lattice results confront Models} 

\author{Massimo D'Elia}
\address{Dipartimento di Fisica dell'Universit\`a 
di Genova and INFN, I-16146, Genova, Italy \\
E-mail : delia@ge.infn.it}
\author{Maria Paola Lombardo}
\address{INFN-Laboratori Nazionali di Frascati, 
I-00044, Frascati(RM), Italy\\
E-mail : lombardo@lnf.infn.it}

\maketitle

\abstracts{ We show that
lattice results  on four flavor QCD at nonzero temperature
and baryon density compare well
with the hadron resonance gas model up to $T \simeq 0.95 T_c$, 
and  approach a free field behaviour  with
a reduced effective number of flavor for $T \ge  1.5 T_c$;
chiral symmetry and confinement are interrelated, and
we note analogies between the critical line of QCD and that
of simple models with the same global symmetries.}
\section{QCD Thermodynamics and Imaginary $\mu_B$}
In the last four years a few 
lattice techniques proven successful in QCD thermodynamics for $\mu_B/T < 1.$
\cite{D'Elia:2002gd,D'Elia:2004at,uno,due,FoPh,tre,susc}.
While waiting for final results in  the scaling limit and with physical
values of the parameters, it is very useful to contrast and compare
current lattice results with model calculations
and perturbative studies. 

The imaginary chemical potential approach\cite{Lombardo:1999cz,Hart:2000ef,FoPh,D'Elia:2002gd,D'Elia:2004at,Giudice:2004se}to QCD thermodynamics 
seems to be ideally suited for the interpretation 
and comparison with analytic results. In the following  we review our
results \cite{D'Elia:2002gd,D'Elia:2004at} from this perspective. 

QCD at finite quark chemical potential 
$\mu$ can be simulated with ordinary methods
when $\mu$ is purely imaginary. If one were able to determine 
thermodynamic observables with
infinite accuracy, standard complex analysis arguments would 
guarantee that the result will be valid within the 
entire analytic domain, i.e. everywhere away from phase 
transitions. In practical numerical work one has to take into account
two  sources of errors: first, the analytic form of the fitting
function is not known a priori; second, even if it were so,
one has to deal with numerical errors on the coefficients. Cross
checks among different analysis and  guidance from models 
are thus most useful.
\begin{figure}[b]
{\epsfig{file= 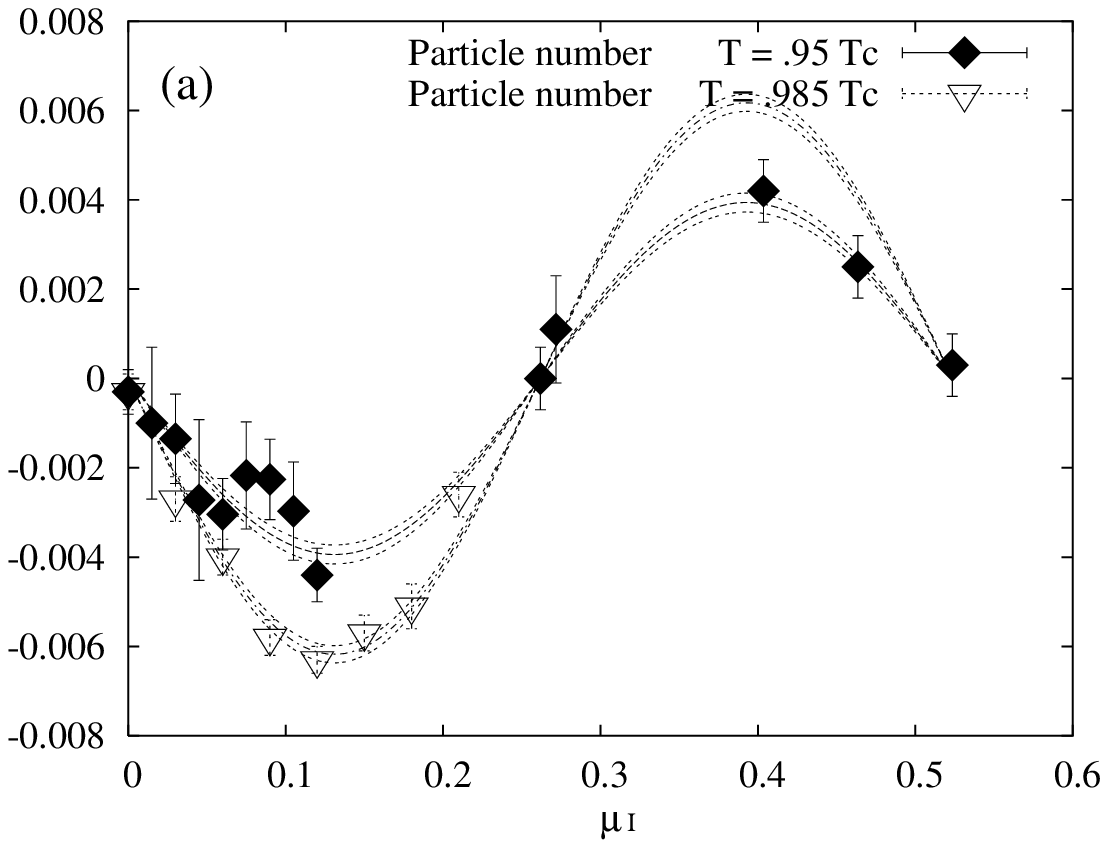, width= 2.2 in}}
{\epsfig{file= 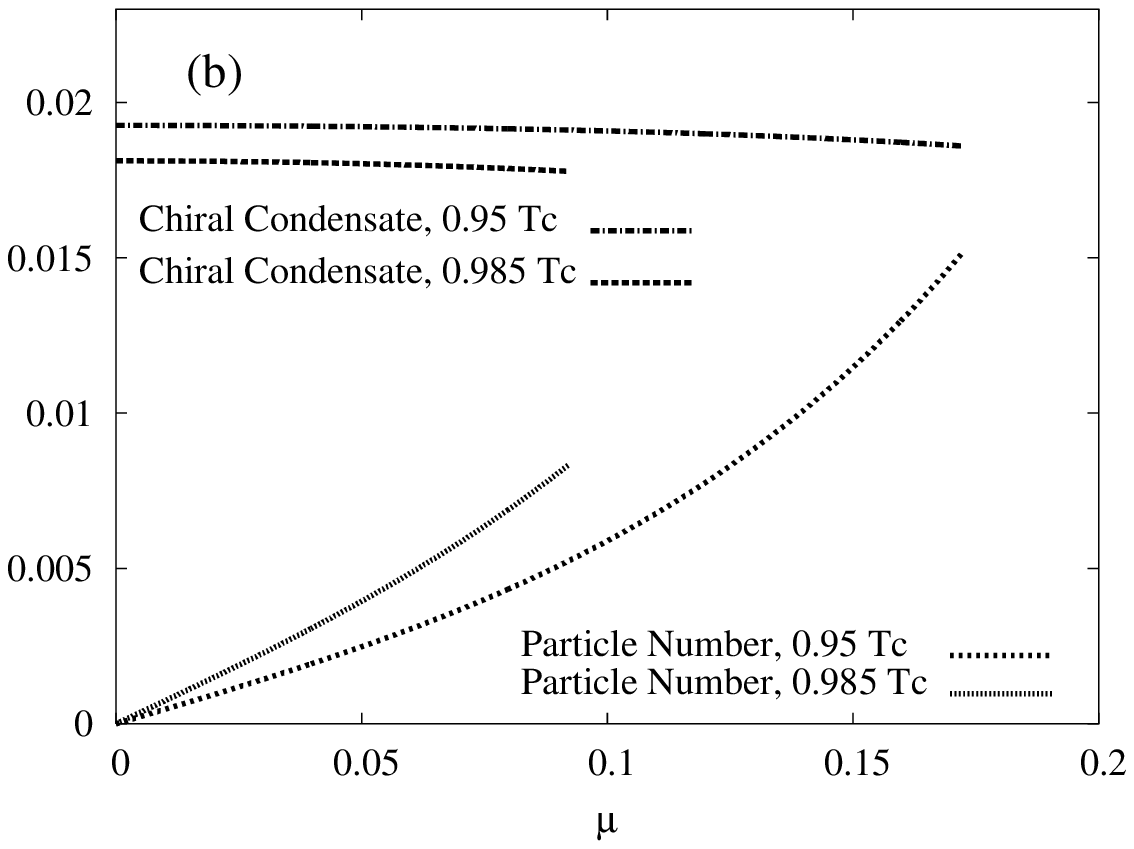, width= 2.2 in }}
\caption{Hadronic Phase:
(a) One Fourier coefficient fit to the particle number, showing
that the Hadron Resonance Model is adequate to describe this data.
(b) Compilation of the results for the chiral
condensate and the particle number as a function
of real chemical potential: the lines are cut in correspondence
with $\mu_c$, showing the first order character
of the phase transition (inferred from the 
chiral condensate) and the critical density.
See our extended writeups for error analysis and details.}
\end{figure}
Results from an imaginary 
$\mu$ have been obtained for the critical line of the two, three and
two plus one flavor model \cite{FoPh}, as well as for 
four flavor \cite{D'Elia:2002gd}.
Thermodynamics results -- order parameter, pressure, number
density -- were obtained for the four flavor model
\cite{D'Elia:2004at} . The pressure was defined by
introducing an integral method at fixed temperature, and
the mass dependence was estimated via the Maxwell relations
\cite{Kogut:1983ia,D'Elia:2004at,susc}.

\section{The Critical Line, Chiral Symmetry and Confinement}
We used the exact results available for the critical line of
the Gross Neveu model, and for Random Matrix Models in the
appropriate universality class  to show
that a second order polynomial in $T$ and $\mu$
 approximates well the critical
line over a large $\mu$ interval\cite{D'Elia:2002gd}.
The nature of the chiral transition
itself can be studied with a great accuracy :
the correlation between $<\bar \psi \psi>$ and Polyakov loop
was observed at several values $\mu_I$, either by studying the results
as a function of $\beta$, and by following the 
Monte Carlo histories directly at $\beta_c(\mu_I)$ 
\cite{D'Elia:2002gd,D'Elia:2004at}.
\section{The Hadronic Phase}
 The grand canonical partition function of the
Hadron Resonance Gas model\cite{Karsch:2003zq} 
has a simple hyperbolic cosine behaviour. This
can be framed in our discussion of the phase diagram in the 
temperature-imaginary 
chemical potential plane which suggests
to use Fourier analysis in this region, as observables are periodic
and continuous there\cite{D'Elia:2002gd}.

For observables which are even ($O_e$) or odd ($O_o$) under
$\mu \to -\mu$    the analytic continuation
to real chemical potential of the Fourier series read
$O_e[o](\mu_I, N_t)   =  \sum_n  a_{F}^{(n)} \cosh [\sinh](
n N_t N_c \mu_I)$.
In our  Fourier analysis of the chiral condensate
\cite{D'Elia:2002gd}
 and of the number density\cite{D'Elia:2004at} - 
even and odd observables, respectively -  
we limited ourselves to $n=0,1,2$ and we assessed the validity
of the fits via both the value of the $\chi^2/{\rm d.o.f.}$ and the stability
of  $a_{F}^{(0)}$ and $a_{F}^{(1)}$ given by one and two cosine 
[sine] fits:
we found that one cosine [sine] fit  describes reasonably well
the data up to $T \simeq 0.985T_c$ (see Fig.1a); 
further terms in the expansion 
did not modify much the value of
the first coefficients and does not particularly  
improve the $\chi^2/{\rm d.o.f.}$. This means that
our data are well approximated by the hadron resonance gas prediction
$\Delta P \propto (\cosh (\mu_B/T) - 1)$
in the broken phase up to $T \simeq 0.985 T_c$. The analysis of the
corrections requires better precision.

The analytic continuation (Fig. 1b) of any 
observable $O$ is valid within the analyticity domain, i.e.
till $\mu < \mu_c(T)$, where $\mu_c(T)$ has to be measured independently.
The value of the analytic continuation of $O$ at
$\mu_c$, $O(\mu_c)$, defines its critical
value. When $O$ is an order parameter which is zero in the quark gluon
plasma phase,  the calculation of $O(\mu_c)$
allows the identification of the order of the phase transition:
first, when $O(\mu_c) \ne 0$, second, when $O(\mu_c) = 0$
\cite{D'Elia:2002gd,D'Elia:2004at}. 
\section{The Hot Phase}
The behaviour of the number density (Fig. 2a) approaches the lattice
Stephan-Boltzmann prediction, with some residual deviation.
We parametrise the deviation  from a free field behavior as 
\cite{Szabo:2003kg,Letessier:2003uj}
\begin{equation}
\Delta P (T, \mu)  =  f(T, \mu) P^L_{free}(T, \mu) 
\end{equation}
where $P^L_{free}(T, \mu)$ is the lattice free result for the pressure.
For instance, in the discussion of Ref. \cite {Letessier:2003uj}
\begin{equation}
f(T, \mu) = 2(1 - 2 \alpha_s/ \pi)
\end{equation}
and the crucial point was that $\alpha_s$ is $\mu$ dependent.

We can search for such a non trivial prefactor $f(T, \mu)$ by taking 
the ratio between the numerical data and the lattice
free field result $ n^L_{free}(\mu_I)$  at imaginary chemical potential:
\begin{equation} 
R(T, \mu_I) = \frac{ n(T,  \mu_I)}{n^L_{free}( \mu_I)}
\end{equation}
A non-trivial (i.e.
not a constant) $R(T, \mu_I)$ would indicate a non-trivial 
$f(T, \mu)$.

In Fig. 2b  we plot $R(T, \mu_I)$ 
versus $\mu_I/T$: the results for $T \ge 1.5 T_c$ seem 
consistent with a free lattice
gas, with an fixed effective number of flavors $N^{eff}_f(T)/ 4 =  R(T) $:
$N^{eff}_f=  0.92 \times 4$ for $T=3.5 T_c$,  and 
$N^{eff}_f = 0.89 \times 4$ for $T = 1.5 T_c$.
\begin{wrapfigure}{r}{6.0 truecm}
{\epsfig{file= 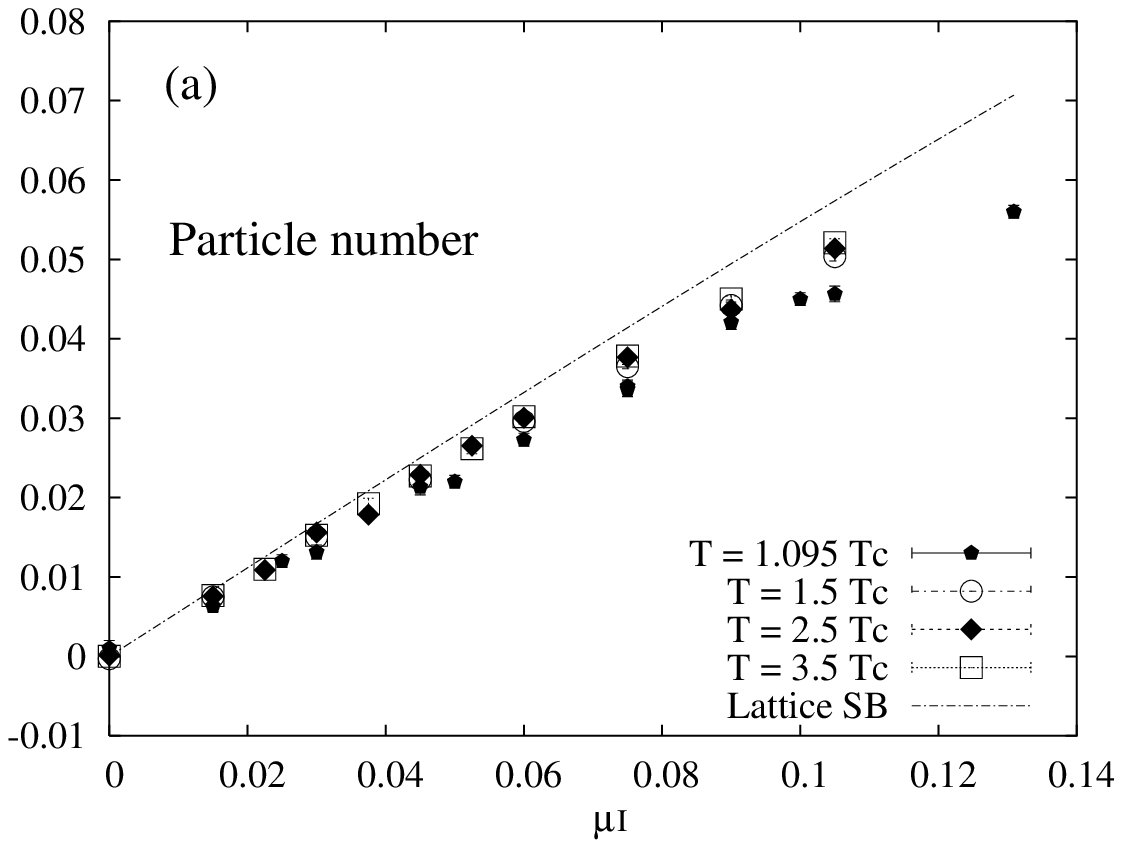, width= 6.0 truecm}}
{\epsfig{file= 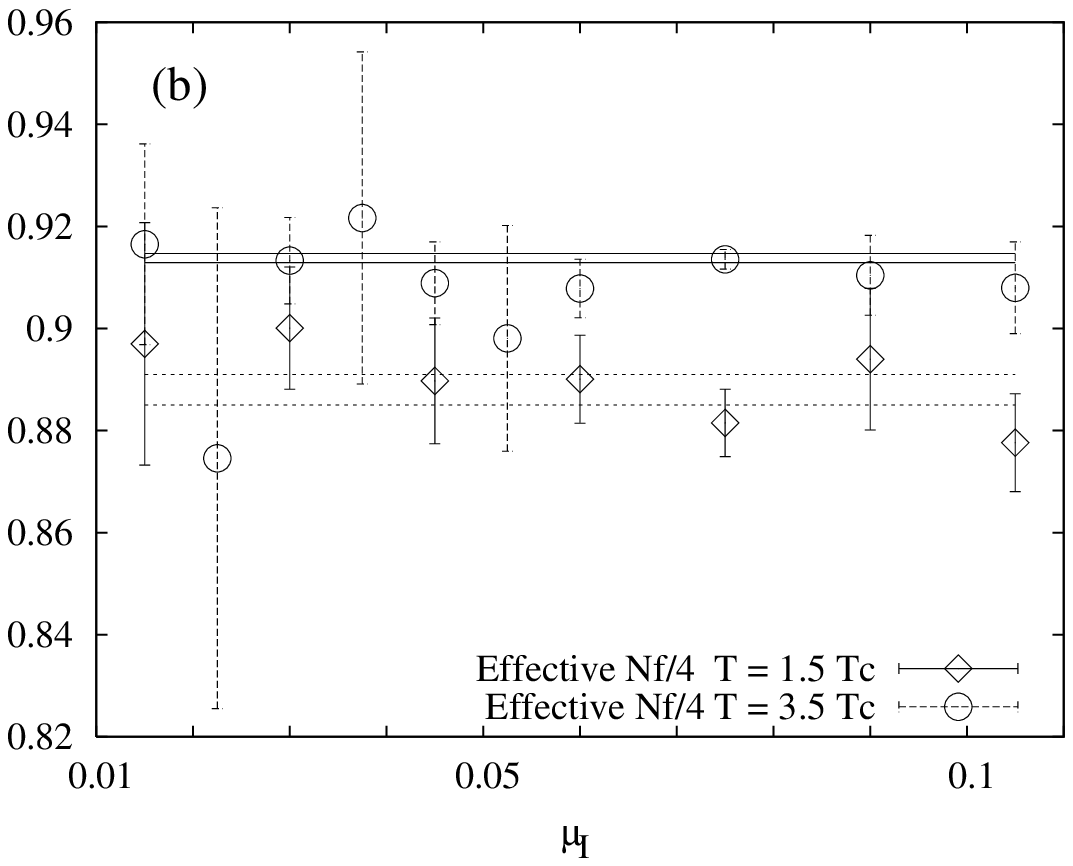, width= 6.0 truecm}}
\caption{Hot Phase: (a) Particle number aproaching the free field behaviour
(b) Ratio of the lattice results to the lattice free field:
 the deviation from free field can be described
by an effective number of flavor $N < 4$  for $T \ge 1.5 T_c$}
\end{wrapfigure}
\section{Lattice vs. Models: \\ Open Issues}
We give here a partial list of 
results together with related questions
which calls for a more detailed understanding of the
interrelations of analytic and numerical results.

 The critical line is similar to the one of model theories
with the same global symmetries: $T/T_c^2 = 1 - 0.0021(2) (\mu/T)^2$
for four flavor, and similarly for two and three flavors 
\cite{uno,FoPh,tre}. Can we understand the flavor dependence and  
the (small) coefficient of  $(\mu/T)^2$ from these model?

We have observed $T^{chiral}_c(\mu) = T^{deconfining}_c(\mu)$
Can  effective Lagrangians\cite{Mocsy:2003qw} 
account for this observation ?

In the hadronic phase \\$\Delta P \simeq k  (1- \cosh (\mu_B/T))$.
The corrections are not completely negligible, and can be used to
estimate the magnitude of further terms in the expansion. 
Can these corrections be modeled in some simple, intuitive way?

 For $T\ge 1.5 T_c$ the results are
compatible with lattice Stefan Boltzmann with
an active fixed  number of flavor $4 \times 0.92$ for 
T=3.5 $T_c$ and $4 \times 0.89$ for
$T = 2.5 T_c$. 
These finite density corrections appear to leave unchanged 
the free field structure,
even at moderately low temperatures. 
How can we understand this? 

Between $T_c$ and 1.5 $T_c$, can we interpret
the deviations from this simple behaviour by use
of a rigorous 
perturbative analysis \cite{Vuorinen},
and / or in the framework of a
strongly interacting quark gluon plasma
\cite{SQGP}?
\section*{Acknowledgments}
This work was partially supported by MIUR.
MPL wishes to thank the Institute for Nuclear Theory at the University
of Washington for its hospitality and the Department of Energy
for partial support during the completion of this work.


\begin{thebibliography}{0}
\bibitem{D'Elia:2002gd}
M.~D'Elia and M.~P.~Lombardo,
Phys.\ Rev.\ D {\bf 67}, 014505 (2003).
\bibitem{D'Elia:2004at}
M.~D'Elia and M.~P.~Lombardo,
arXiv:hep-lat/0406012.
\bibitem{uno}
Z.~Fodor and S.~D.~Katz,
Phys.\ Lett.\ B {\bf 534}, 87 (2002); 
JHEP {\bf 0404}, 50 (2004); \\ 
Z.~Fodor, S.~D.~Katz and K.~K.~Szabo,
Phys.\ Lett.\ B {\bf 568}, 73 (2003); \\
F.~Csikor {\em et al.} JHEP {\bf 0405} (2004) 046;\\
S.~D.~Katz, Nucl.\ Phys.\ Proc.\ Suppl.\  {\bf 129}, 60 (2004).
\bibitem{due}
Ph.~de Forcrand {\em et al.},
Nucl.\ Phys.\ Proc.\ Suppl.\  {\bf 119}, 
541 (2003)
\bibitem{FoPh}
Ph.~de Forcrand and O.~Philipsen,
Nucl.\ Phys.\ B {\bf 642}, 290 (2002);\\
Nucl.\ Phys.\ B {\bf 673}, 170 (2003).
\bibitem{tre}
 C.~R.~Allton {\it et al.},
Phys.\ Rev.\ D {\bf 66}, 074507 (2002); \\
 Phys.\ Rev.\ D {\bf 68}, 014507 (2003).
\bibitem{susc} R.~Gavai, S.~Gupta and R.~Roy,
Prog.\ Theor.\ Phys. \ Suppl.{\bf 153}, 270 (2004).
\bibitem{Lombardo:1999cz}
M.~P.~Lombardo,
Nucl.\ Phys.\ Proc.\ Suppl.\  {\bf 83}, 375 (2000).
\bibitem{Hart:2000ef}
A.~Hart, M.~Laine and O.~Philipsen,
Phys.\ Lett.\ B {\bf 505}, 141 (2001).
\bibitem{Giudice:2004se}
P.~Giudice and A.~Papa,
Phys.\ Rev.\ D {\bf 69}, 094509 (2004).
\bibitem{Kogut:1983ia}
J.~B.~Kogut {et al.}
Nucl.\ Phys.\ B {\bf 225}, 93 (1983).
\bibitem{Karsch:2003zq}
F.~Karsch, K.~Redlich and A.~Tawfik,
Phys.\ Lett.\ B {\bf 571}, 67 (2003).
\bibitem{Szabo:2003kg}
K.~K.~Szabo and A.~I.~Toth, JHEP {\bf 0306}, 008 (2003).
\bibitem{Letessier:2003uj}
J.~Letessier and J.~Rafelski,
Phys.\ Rev.\ C {\bf 67}, 031902 (2003).
\bibitem{Vuorinen}
A.~Vuorinen,
arXiv:hep-ph/0402242; \\ A.~Vuorinen,
Phys.\ Rev.\ D {\bf 68}, 054017 (2003);\\
A.~Ipp, A.~Rebhan and A.~Vuorinen,
Phys.\ Rev.\ D {\bf 69}, 077901 (2004).
\bibitem{Mocsy:2003qw} 
A.~Mocsy, F.~Sannino and K.~Tuominen,
Phys.\ Rev.\ Lett.\  {\bf 92}, 182302 (2004)
\bibitem{SQGP} See e.g.the WEB page of the RBRC Workshop\\ 
{\em New Discoveries at RHIC:}{\em the current case for the strongly interactive 
QGP}\\  {\tt http://www.bnl.gov/riken/May14-152004workshop.htm}

\end{thebibliography}
\end{document}